\documentclass[twocolumn]{aastex62}

\emergencystretch=\maxdimen
\graphicspath{{./}{figures/}}

\usepackage[none]{hyphenat}

\newcommand{\RIT}{\affiliation{Center for Computational Relativity and Gravitation, Rochester Institute of Technology, Rochester, New York 14623, USA}}

\date{\today}

\shorttitle{IMBH trigger in LIGO}
\shortauthors{Udall et al.}

\begin{document}

\title{Inferring Parameters of GW170502: The Loudest Intermediate-mass Black Hole Trigger in LIGO's O1/O2 data}

\author{Richard Udall}
\affiliation{Center for Relativistic Astrophysics, School of Physics, Georgia Institute of Technology, 837 State St., Atlanta, GA, USA - 30363 }

\author[0000-0003-1007-8912]{Karan Jani}
\affiliation{Center for Relativistic Astrophysics, School of Physics, Georgia Institute of Technology, 837 State St., Atlanta, GA, USA - 30363 }
\affil{Department Physics and Astronomy, Vanderbilt University, 2301 Vanderbilt Place, Nashville, TN, 37235, USA}

\author{Jacob Lange}
\RIT

\author{Richard O'Shaughnessy}
\RIT

\author{James Clark}
\affiliation{Center for Relativistic Astrophysics, School of Physics, Georgia Institute of Technology, 837 State St., Atlanta, GA, USA - 30363 }

\author{Laura Cadonati}
\affiliation{Center for Relativistic Astrophysics, School of Physics, Georgia Institute of Technology, 837 State St., Atlanta, GA, USA - 30363 }

\author{Deirdre Shoemaker}
\affiliation{Center for Relativistic Astrophysics, School of Physics, Georgia Institute of Technology, 837 State St., Atlanta, GA, USA - 30363 }

\author{Kelly Holley-Bockelmann}
\affil{Department Physics and Astronomy, Vanderbilt University, 2301 Vanderbilt Place, Nashville, TN, 37235, USA}

\begin{abstract}
Gravitational wave (GW) measurements provide the most robust constraints of the mass of astrophysical black holes. Using state-of-the-art GW signal models and a unique parameter estimation technique, we infer the source parameters of the loudest marginal 
trigger, GW170502, found by LIGO from 2015 to 2017. If this trigger is assumed to be a binary black hole merger, we find it corresponds to a total mass in the source frame of $157^{+55}_{-41}~\rm{M}_\odot$ at redshift  $z=1.37^{+0.93}_{-0.64}$. 
The primary and secondary black hole masses are constrained to $94^{+44}_{-28}~\rm{M}_{\odot}$ and $62^{+30}_{-25}~\rm{M}_{\odot}$ respectively, with 90\% confidence. Across all signal models, we find $\gtrsim 70\%$ probability for the effective spin parameter $\chi_\mathrm{eff}>0.1$. Furthermore, we find that the inclusion of higher-order modes in the analysis narrows the confidence region for the primary black hole mass by 10\%, however, the evidence for these modes in the data remains negligible. 
The techniques outlined in this study could lead to robust inference of the physical parameters for all intermediate-mass black hole binary candidates $(\gtrsim100~\mathrm{M}_\odot)$ in the current GW network. 

\end{abstract}

\keywords{black hole physics --- gravitational waves ---  stars: black hole}

\section{Introduction} \label{sec:intro}
Gravitational-wave (GW) detectors have begun a unique survey of the transient sky, enabling a census of coalescing binary black holes (BBHs) in the local universe. The Advanced LIGO \citep{aLIGO2}/Virgo \citep{aVirgo2} network is most sensitive to BBH mergers with binary masses of order ~$\sim400~\mathrm{M}_\odot$ in the detector frame \citep{ObservingScenarios, jani2019detectability}. No event with binary mass above $\sim 100~\mathrm{M}_\odot$ was found in the first and second observing runs of these detectors (O1/O2, 2015-2017, $\sim 166$ days of total data) with the probability of astrophysical origin over 50\% \citep{O2Catalog, O2IMBH}. However, the loudest {\it marginal} candidate reported across O1/O2 data among all LIGO-Virgo compact binary searches occurred on May 2nd, 2017.

\begin{table*}[ht!]
    \centering
 \begin{tabular*}{\textwidth}{l @{\extracolsep{\fill}} llll}
 \hline \hline
    Waveform Model & \multicolumn{2}{ c }{\texttt{NRHybSur3dq8} (Aligned Spins)}  & \multicolumn{2}{c }{\texttt{NRSur7dq4} (Precessing Spins)}   \\ \hline
    Radiated Modes ($\ell, m$)  & $\ell \leq 4$, $(5,5)$ & $\ell=2$ & $\ell \leq 4$ & $\ell=2$ \\\hline
    Primary BH mass, $m_1^\mathrm{src}~[\mathrm{M}_\odot]$ & $94^{+44}_{-28}$ & $104^{+56}_{-35}~$& $96^{+45}_{-30}~$&$112^{+66}_{-40}~$ \\
    Secondary BH mass, $m_2^\mathrm{src}~[\mathrm{M}_\odot]$ & $62^{+30}_{-25}~$&$63^{+32}_{-27}~$ &$62^{+34}_{-24}~$ & $65^{+36}_{-24}~$\\
    Total mass, $M^\mathrm{src}_\mathrm{tot}$ & $157^{+55}_{-41}~$&$169^{+60}_{-48}~$ & $159^{+61}_{-44}~$& $179^{+76}_{-53}~$\\
    Chirp mass, $\mathcal{M}_c^\mathrm{src}$ &$65^{+24}_{-17}~$ & $69^{+25}_{-20}~$& $66^{+27}_{-18}~$& $73^{+30}_{-21}~$\\
    Mass ratio, $q = m_2/m_1$ &$0.69^{+0.27}_{-0.37}$ &$0.64^{+0.32}_{-0.36}$ &$0.68^{+0.28}_{-0.35}$ & $0.60^{+0.34}_{-0.31}$\\
    Effective inspiral spin, $\chi_\mathrm{eff}$ & $0.49^{+0.31}_{-0.63}$& $0.49^{+0.32}_{-0.62}$&$0.28^{+0.35}_{-0.49}$ &$0.26^{+0.35}_{-0.45}$ \\
    Effective precession spin, $\chi_p$  & - & - & $0.60^{+0.24}_{-0.32}$ & $0.62^{+0.23}_{-0.32}$\\
    Redshift, $z$ & $1.37^{+0.93}_{-0.64}$ & $1.20^{+0.93}_{-0.59}$& $1.29^{+0.87}_{-0.64}$& $1.02^{+0.84}_{-0.57}$\\
    
\end{tabular*}
\caption{Parameters of GW170502 for the two waveform models and different combinations of modes discussed in this study.}
    \label{tab:1}
\end{table*}
Henceforth known as GW170502 for consistency in the nomenclature adapted in the literature for candidates of similar statistical significance (such as GW170817A \citep{2019arXiv191009528Z} and GW151205 \citep{2019arXiv191005331N}), this trigger was found with a false alarm rate of $0.34~\mathrm{yr}^{-1}$ by a transient burst search developed especially for detecting mergers of intermediate mass black holes (IMBHs)\footnote{Here, we define IMBH binaries to be between $100-1000 \, M_\odot$, regardless of formation mechanism. We acknowledge the potential difference with astrophysical definitions of IMBHs} in LIGO-Virgo detectors \citep{Klimenko:2015ypf}. Follow up parameter estimation of this trigger suggested a source frame chirp mass of {${\sim70}~\rm{M}_{\odot}$}. The candidate had a signal-to-noise ratio $S/N\sim6$, less than the standard detection threshold of $S/N \geq 8$. Due to the very high detector frame mass, the signal was essentially a GW burst. The two additional matched-filtering algorithms \citep{PhysRevD.95.042001, Usman_2016} that contributed to IMBH search in O1/O2 registered this candidate at much lower significance (see Appendix D of \cite{O2IMBH} for discussion). 

Along with GW170817A \citep{2019arXiv191009528Z} and GW151205 \citep{2019arXiv191005331N}, the candidate GW170502 adds to an emerging population of high mass binary analyzed in the open data era. All of these open-data events have similar statistical significance, i.e., they will be considered {\it marginal} candidates in comparison to the confirmed LIGO-Virgo events. However, no such triggers in GW astronomy are unique, meaning if we see it once, we will see something similar again; and, therefore the purpose of this study is not to make new statements on the detection confidence of GW170502 itself, 
or its implications for the BH population. The probability of astrophysical origin for such triggers is subject to change with better sensitivity of LIGO/Virgo detectors. The goal here is to demonstrate through GW170502 a new machinery that allows rapid inference of the properties of IMBH candidates $(\gtrsim 100~\mathrm{M}_\odot)$ in the current GW network.

In this study, we reanalyze O1/O2 data at the instance of GW170502 to determine its potential source parameters with two new state-of-the-art tools: (i) a rapid Bayesian inference algorithm \citep{2017arXiv170509833L, 2018arXiv180510457L}, and (ii) numerical relativity simulation-based waveform models that includes radiated higher order modes (HOMs) $(\ell,|m|)$ beyond the dominant $(2,2)$ mode \citep{2019PhRvD..99f4045V, 2019arXiv190509300V}. These HOMs are critical for analysis of such massive BBH mergers \citep{PhysRevD.97.024016}. We find that our reanalysis tightens the constraint on the total-mass of this trigger, as well as suggests a positive effective spin. This is consistent with the hierarchical formation channels for binaries in this mass range \citep{2019PhRvD.100d3027R, yang2019hierarchical, Gayathri_2020}.

\begin{figure*}[t!]
\centering
\includegraphics[scale=0.54,trim = {100 -10 0 0}]{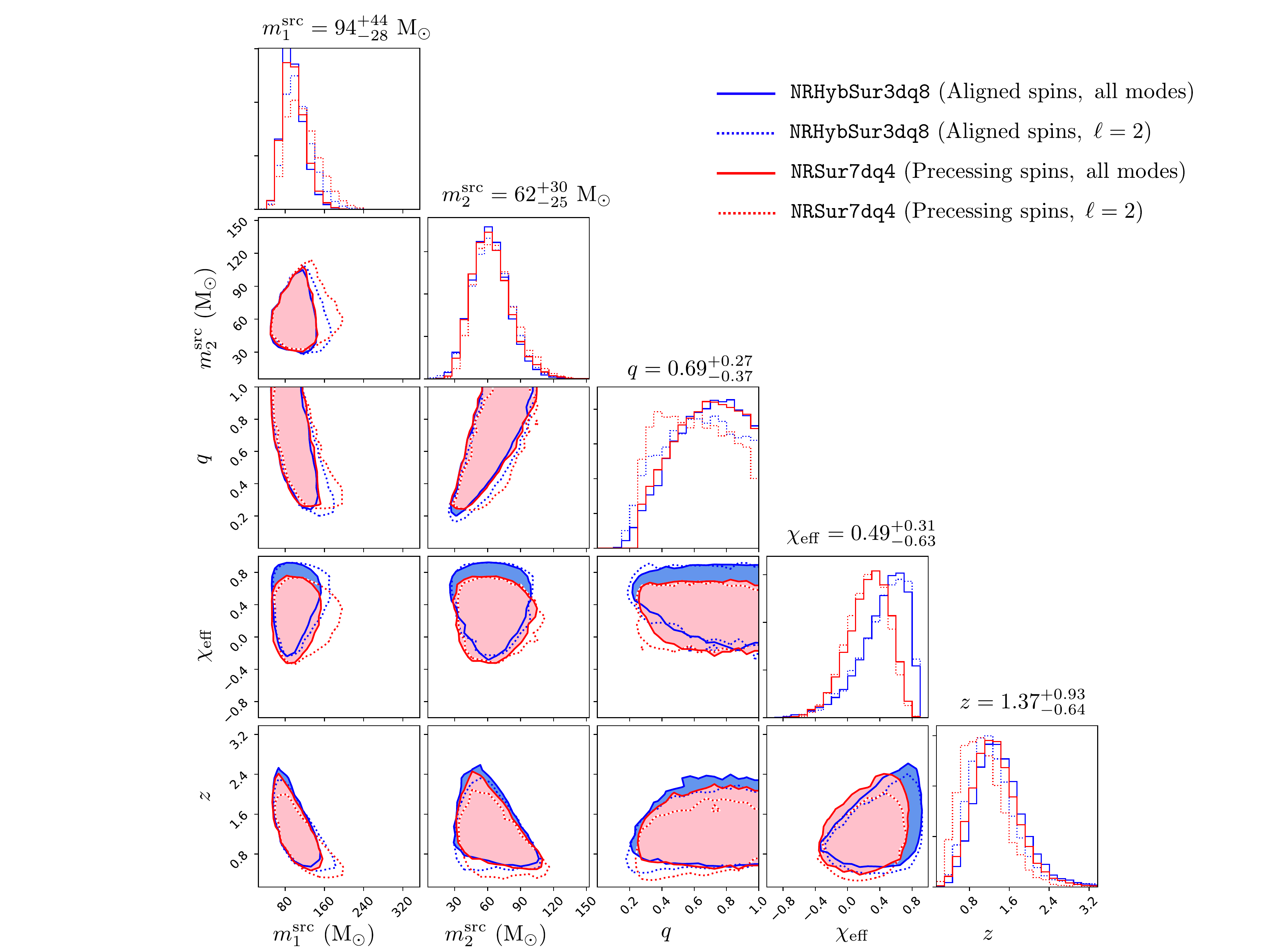}
\caption{Marginalized posteriors for GW170502 using \texttt{RIFT}. Two-dimensional contours enclose $90\%$ of the distribution. The two colors refer the two waveform models \texttt{NRHybSur3dq8}  (blue) and \texttt{NRSur7dq4} (red). The solid lines refer to results for including all the higher-order modes in the waveform (\texttt{NRHybSur3dq8} with $\ell\leq 5$ and \texttt{NRSur7dq4} with $\ell\leq 4$), the dotted lines represent models restricted to $\ell =2$. The numbers quoted above each column are the median, with the 90\% interval obtained from \texttt{NRHybSur3dq8} using all the available higher-order modes.
}
\label{fig1}
\end{figure*}


\begin{figure}[t!]
\centering
\includegraphics[scale=0.33,trim = {140 0 0 30}]{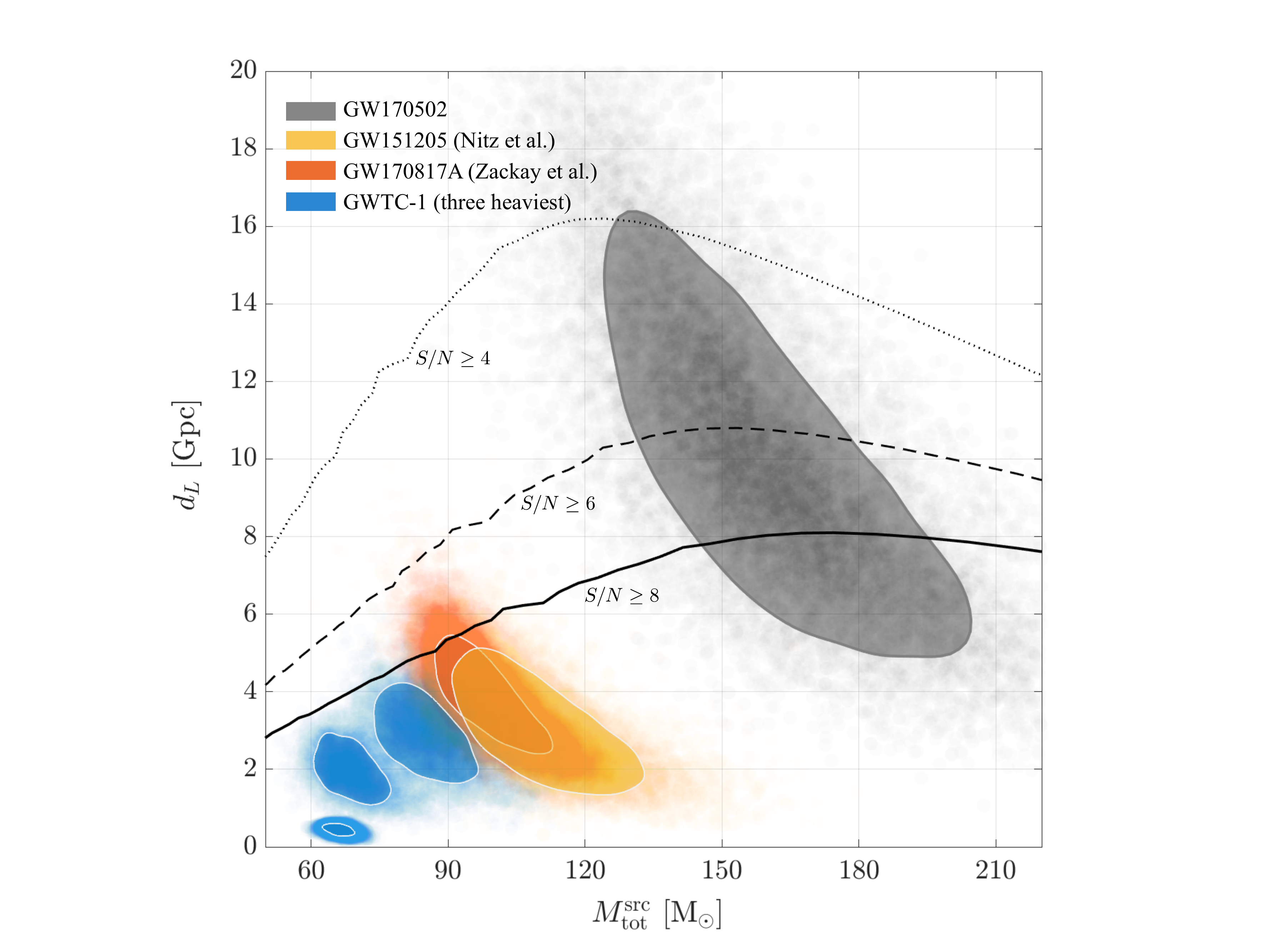}
\caption{Six heaviest BBH mergers reported from the O1/O2 runs of Advanced LIGO/Virgo (2015-2017). The horizontal axis is the total mass in the source frame and vertical axis is their corresponding luminosity distance. The contours refer to 90\% confidence intervals and the transparent dots show the spread of the posterior sample. With the black contour, we show the constraints on GW170502 using the \texttt{NRHybSur3dq8} model with all the available higher-order modes. The blue contours show the three heaviest confirmed BBH mergers -- GW170729, GW170823, and GW150914 -- as reported in GWTC-1 \citep{O2Catalog}. In the orange and yellow contours, we show candidate GW170817A and GW151205 found by independent teams \citep{2019arXiv191009528Z, 2019arXiv191005331N}. The horizon distances for nonspinning, equal-mass BBHs (black curves) are computed at different S/N for a single detector Advanced LIGO sensitivity during O2.} 
\label{fig2}
\end{figure}

\section{Methodology} \label{sec:methods}
To infer binary parameters via Bayesian inference, we use an algorithm called \texttt{RIFT} (Rapid parameter Inference
via Iterative Fitting, \cite{2017arXiv170509833L, 2018arXiv180510457L}). This algorithm iteratively constructs and refines an approximation to the marginal likelihood over the extrinsic ($\theta:=$ distance, sky location, inclination) and intrinsic ($\lambda:=$ BH masses, spins) binary parameters,
\begin{equation} \label{eq:lnLmarg}
 {\cal L}_{\rm marg}\equiv\int  {\cal L}(\lambda, \theta )~p(\theta )~\mathrm{d}\theta 
 \end{equation}
\texttt{RIFT's} structure and organization is particularly well-suited to analyzing GW signal models with HOMs. Internally, \texttt{RIFT} decomposes the outgoing radiation into spherical harmonics to compress and accelerate the likelihood. In general, models with HOMs are computationally intense for standard parameter estimation tools such as \texttt{LALInference} \citep{2015PhRvD..91d2003V}, taking days to a week for analyzing comparable high-mass sources with a single approximation. For heavy mass systems like GW170502, it is crucial to always use signal models with HOM \citep{PhysRevD.97.024016,PhysRevD.97.064027}. As a result, \texttt{RIFT} stands out as an important tool for studying all potential IMBH candidates in the current epoch of GW astronomy.


\texttt{RIFT}, as with all parameter estimation methods, requires waveform models to interpret the GW signals from BBH coalescence. Here, we use two waveform models: (i) \texttt{NRHybSur3dq8}, which is our preferred model in this study \citep{2019PhRvD..99f4045V}, and (ii) \texttt{NRSur7dq4}, which we utilized for additional checks (i.e. information about precessing spins) \citep{2019arXiv190509300V}. While both models are tuned directly to numerical relativity simulations, the former is the only hybrid model. Due to this source having such a high mass, the duration length of the waveforms is irrelevant. Model (i) is valid up to mass-ratio $m_2/m_1 \geq 1/8$ and for BBHs with aligned spins $|\chi_{1z,2z}|\leq 0.8$ (in dimensionless units) but extrapolates well to $|\chi_{1z,2z}|\leq 0.9$, while Model (ii) is valid up to mass-ratio $m_2/m_1 \geq 1/4$ and for BBHs with generic spin orientation that captures the spin-orbit precession with the same spin magnitude restrictions. In context of GW170502, the specific advantage of using these two models is the inclusion of the radiated modes up to $\ell\leq4$ and $(\ell,|m|)=(5,5)$. 

We use both these waveform models to conduct a full parameter estimation of GW170502 with the \texttt{RIFT} algorithm. To test the impact of HOMs, we conduct a separate estimation only including the dominant $\ell=2$ modes using 
both the models (see Fig. \ref{fig1} and Table \ref{tab:1}). 
We adopt the conventional mass and distance priors for Advanced LIGO/Virgo data analysis: a uniform mass density in the detector frame and uniform in the cube of the luminosity distance. For the aligned spin analyses, we adopt a uniform prior for $\chi_{i,z} \in [-0.9, 0.9]$, component of BH spins aligned with angular momentum, and assume there is no in-plane spin components. For the precessing analysis, we adopt a spin prior where the spin vectors are uniformly distributed within the unit sphere. 

\section{Results \& Discussions} \label{sec:results}
\paragraph{Masses} By using all the available HOMs in the \texttt{NRHybSur3dq8} waveform model, we find that GW170502 corresponds to a total binary mass in the source frame of $M_\mathrm{tot}^{\mathrm{src} }= 157^{+55}_{-41}~\rm{M}_{\odot}$ with 90\% confidence interval. This makes the trigger heavier than all previous sources in the first and second observing runs of Advanced LIGO (see Fig. \ref{fig2}). 
The corresponding redshift and luminosity distance is constrained to $z=1.37^{+0.93}_{-0.64}$ and $d_L =10^{+8.9}_{-5.4}$~Gpc, which makes it potentially the farthest GW source. The observed trigger, therefore, was strongly redshifted, and the detector frame mass was about $\sim 3$ times heavier.  The primary BH mass in the source frame was constrained to $m_1^{\mathrm{src} }=94^{+44}_{-29}~\rm{M}_{\odot}$
and the secondary BH to $m_2^{\mathrm{src} }=62^{+30}_{-25}~\rm{M}_{\odot}$. 
While the constraints on mass ratio are usually not as stringent with such massive binaries, we find that both waveform models put GW170502 at $q=m_2/m_1\gtrsim 1/4$ within 90\% confidence.

At the total mass of GW170502, the sensitivity of Advanced LIGO detectors in O2 at the detection threshold ($S/N=8$) was up to ${\sim}8$~Gpc. Considering this trigger was sub-threshold with a network $S/N\sim 6$, it is reasonable that LIGO Livingston (the more sensitive detector) must have recorded $S/N\gtrsim 6/\sqrt{2} \sim 4$. 
This suggests GW170502 was well within the horizon volume (see Fig. \ref{fig2}).

\paragraph{Spins} For such a heavy binary, only the merger is essentially recorded in the LIGO frequency band. Therefore, the individual BH spins and their evolution remain ill-constrained.
We, therefore, focus on constraining the effective inspiral spins, $\chi_\mathrm{eff}$, the net component of mass-weighted spins projected on the orbital angular momentum axis (for definition, see \cite{Ajith:2009bn}). 
For GW170502, we constrain $\chi_\mathrm{eff} = 0.49^{+0.31}_{-0.63} $ with 90\% confidence. The median and upper-bounds of $\chi_\mathrm{eff}$ are higher than reported earlier for this trigger (see Appendix-D of \cite{O2IMBH}). It is also significantly higher than all the BBH mergers of GWTC-1 (see Table III of \cite{O2Catalog}). For our preferred model \texttt{NRHybSur3dq8}, we find that the Bayes' Factor $(\mathcal{B})$ has a mild preference for spinning BHs over non-spinning ($\log_{10}\mathcal{B} = 0.46$). 

\begin{figure}[t!]
\centering
\includegraphics[scale=0.3,trim = {100 20 0 0}]{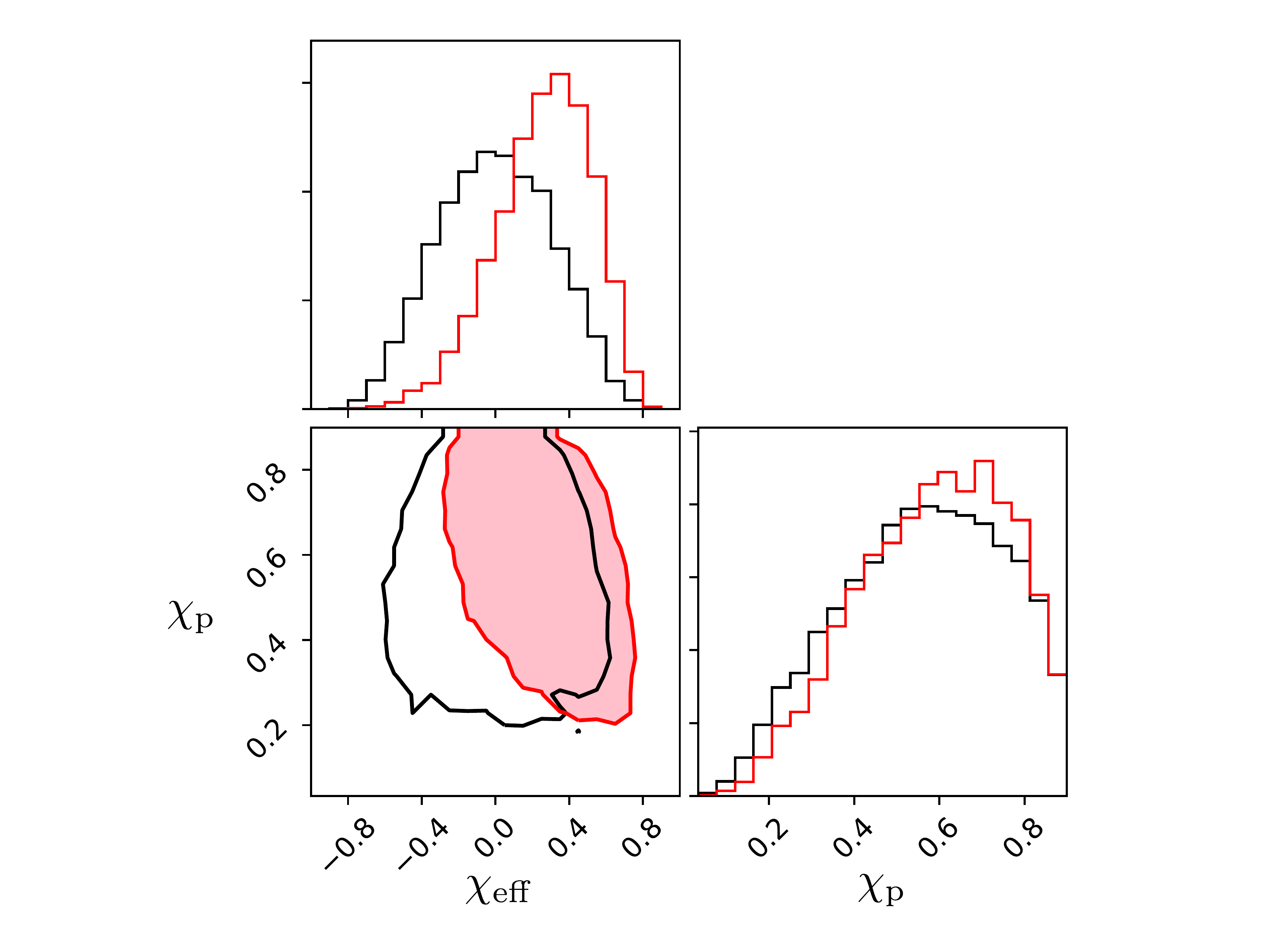}
\caption{Marginalized posteriors of the effective inspiral spin parameter $\chi\mathrm{eff}$ and effective precession spin parameter $\chi_p)$ for GW170502 using \texttt{RIFT}. Two-dimensional contours show $90\%$ intervals for \texttt{NRSur7dq4} model  (red line) and the prior distribution (black line).}
\label{fig3}
\end{figure}

Furthermore, Table \ref{tab:2} shows that in our preferred model, 86.6\% of the posterior lies within the region $0.1 <\chi_\mathrm{eff}$. {When taken with the fraction that lies below $\chi_\mathrm{eff} = -0.1$, we find that over 90\% of the posterior lies outside the region $-0.1<\chi_\mathrm{eff} < 0.1$, thus adding a strong support for a conclusion of non-zero effective inspiral spin.} To investigate if the BH spins of GW170502 have components in the orbital plane, we measured the effective precession spin parameter $\chi_\mathrm{p}$ \citep{PhysRevD.91.024043}. In Fig. \ref{fig3}, we compare the constraints on the effective precession and inspiral parameter ($\chi_p$ vs $\chi_\mathrm{eff}$) for the fully precessing spin model \texttt{NRSur7dq4}. 
Similar to the aligned-spin model, we find $\chi_\mathrm{eff}$ consistently peaks at a positive value even if the model allows generic spin orientations.{We calculate the Bayes' Factor $\mathcal{B}$ between precessing and aligned spin assumptions to be $\log_{10}\mathcal{B} = -0.68$ (disfavoring precession moderately) for the \texttt{NRSur7dq4} model, see Table \ref{tab:3} } However, we gain no new information about $\chi_p$ and the spin-orbit precession of GW170502. 


\begin{table}[t!]
    \centering
    \begin{tabular}{c|cc|cc}
    \hline \hline
        Waveform Model & \multicolumn{2}{c}{\texttt{NRHybSur3dq8}} & \multicolumn{2}{|c}{\texttt{NRSur7dq4}} \\ \hline
        Radiated Modes & $\ell \leq 4$, $(5,5)$ & $\ell = 2$ & $\ell \leq 4$ & $\ell=2$ \\\hline
        $\chi_\mathrm{eff} < -0.1 $ & 6.0\%& 5.9\%& 9.4\%& 9.3\%\\
        $-0.1 < \chi_\mathrm{eff} < 0.1 $ & 7.4\%& 7.8\%& 15.9\%& 18.3\%\\
        $0.1 < \chi_\mathrm{eff} $ & 86.6\%& 86.3\%& 74.7\%& 72.4\%\\
    \end{tabular}
    \caption{Probability from the posterior of each model and mode combination for value of $\chi_\mathrm{eff}$ in the specified bounds. 
    }
    \label{tab:2}
\end{table}

\begin{table}[t!]
    \centering
    \begin{tabular}{c|c|c}
    \hline \hline
        Model & \texttt{NRHybSur3dq8} & \texttt{NRSur7dq4} \\\hline
        $\log_{10} \mathcal{B}$ (HOM) & -0.03 & 0.00 \\
        $\log_{10} \mathcal{B}$ (spinning) & 0.46 & 0.05 \\
        $\log_{10} \mathcal{B}$ (precession) & - & -0.68 \\
    \end{tabular}
    \caption{Bayes' Factors between HOMs vs. non-HOMs, Aligned Spin vs. Zero Spin and Precessing Spin vs. Aligned Spin 
    }
    \label{tab:3}
\end{table}

\begin{figure}[t!]
\centering
\includegraphics[scale=0.24,trim = {0 20 0 0}]{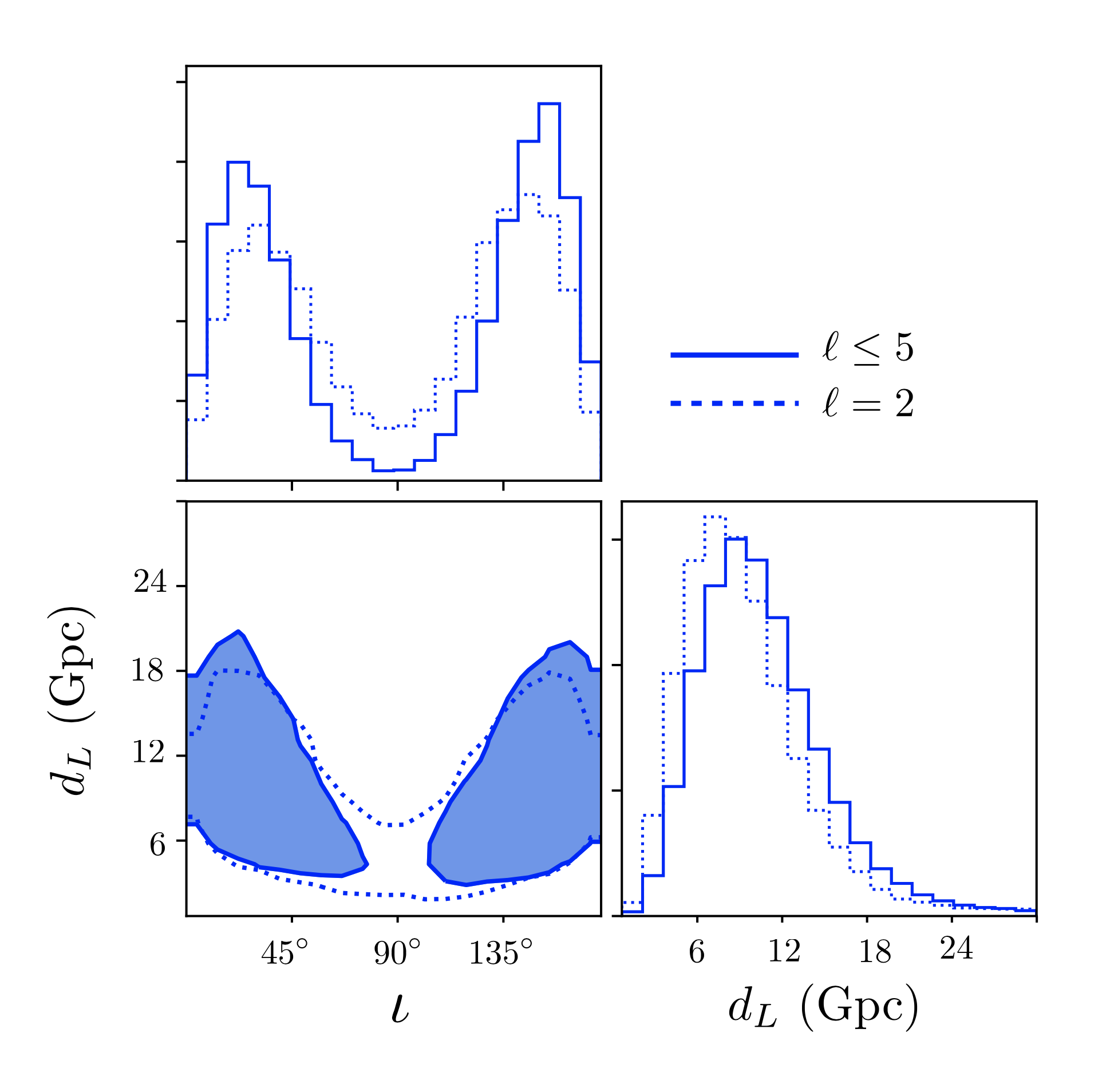}
\caption{Marginalized posteriors of inclination $(\iota)$ and luminosity distance  $(d_L)$ for GW170502 using \texttt{RIFT}. Two-dimensional contours show $90\%$ intervals for \texttt{NRHybSur3dq8} model with (solid line) and without (dotted line) including higher-order modes of gravitational radiation.}
\label{fig4}
\end{figure}

\paragraph{Impact of Higher Order Modes} The dominant mode of gravitational radiation, $(2,2)$, radiates primarily in the direction of net angular momentum, while the (HOMs) carry radiation from off-axis asymmetry. Including the latter in the GW models breaks degeneracy on the extrinsic (particularly inclination angle, $\iota$) as well as intrinsic binary parameters (particularly mass ratio, $q$). For the \texttt{NRHybSur3dq8} model, 
we find that including HOMs narrows our estimate of the inclination angle of GW170502 by ${\sim}25\%$, and more strongly excludes edge-on configurations 
(see Fig. \ref{fig4}). While the $\ell=2$ modes hinted at a low probability for an edge-on orientation $(\iota\sim 90^\circ)$, including $\ell\leq 5$ completely rules it out within 90\% confidence intervals; in fact, this analysis suggests that the binary is close to face-on. Since face-on configurations are more easily detected, the HOMs pushed the distance of GW170502 out by ${\sim}10\%$. This increase in distance (redshift) directly translates into a lower mass for the BHs. For comparison, the median value of the primary BH mass using \texttt{NRHybSur3dq8} is $m_1^\mathrm{src}=94~\rm{M}_{\odot}$ for $\ell\leq5$, while $m_1^\mathrm{src}=104~\rm{M}_{\odot}$ for the $\ell=2$ case.

While the inclusion of HOMs have a significant impact on the posteriors of GW170502, we do not find compelling evidence for their presence in the data. To quantify this, we computed the Bayes' Factor $(\mathcal{B})$ between the $\ell=2$ and $\ell\leq5$ ($\ell\leq4$) case for both the waveform models. As stated in Table \ref{tab:3}, we find $\log_{10} \mathcal{B}$ to be -0.03 and 0.00 for the \texttt{NRHybSur3dq8} and \texttt{NRSur7dq4} signal models respectively.


\paragraph{Conclusions} Our study demonstrates the necessary combination of parameter inference and waveform modeling techniques to constrain IMBH binary mergers.  We apply this machinery to GW170502, the heaviest and loudest BBH trigger found in Advanced LIGO between 2015-2017. Using the most sophisticated GW models, we find that the primary and secondary BH masses of this trigger would correspond to ${\sim} 90~\mathrm{M}_\odot$ and ${\sim 60}~\mathrm{M}_\odot$.
While not reflected in the Bayes' Factors, there is  noticeable shift in the posteriors using the HOMs and $\ell=2$ modes. It narrows the constraints on inclination and distance, thus reducing the uncertainty in BH masses. 

In the next era of GW astronomy, GW170502-like events would be detected in mutliband network of earth-based detectors and space missions such as LISA and the deci-Hz Observatory \citep{2016PhRvL.116w1102S, jani2019detectability, sedda2019missing}. This will increase the detection confidence of such sub-threshold triggers, and substantially improve the constraints on the masses and spins of the two BHs \citep{2016PhRvL.117e1102V, WP-Multiband}. For a future study, we will extend our machinery to explore the impact on parameter estimation of using fully general relativistic simulations of BBH coalescence.

\acknowledgments
We thank Zackay et al. for sharing the posterior samples of 170817A. R.U,  J.C., L.C. and D.S.'s research was funded by the NSF grant PHY 1809572 and 1806580. R.O'S. and J.L.'s research was funded by NSF award PHY 1707965. K.J.'s research was supported by the GRAVITY fellowship at Vanderbilt University.

\paragraph{Computing Resource} The authors are grateful for computational resources used for the parameter estimation runs provided by the the LIGO Lab computing facilities at Caltech, Hanford, and
Livingston, maintained by the California Institute of Technology at Pasadena, California through the LIGO Data Analysis System.

\bibliography{references}

\end{document}